\documentclass[a4paper,10pt]{article}
\usepackage{graphicx}
\usepackage{bm}

\def\eq#1{{Eq.~(\ref{#1})}}
\newcommand{\mdof}{microscopic degrees of freedom}
\newcommand{\LL}{Lanczos-Lovelock}
\def\gu#1#2{{g^{#1#2}}}
\def\gl#1#2{{g_{#1#2}}}

\newcommand{\w}[1]{\bm{#1}}
\newcommand{\el}{\w{\ell}}

\newcommand{\D}{\ensuremath{\nabla}}

\newcommand{\etal}{{\it et al\/}\ }

\title{Lessons  from Classical Gravity about the\\
  Quantum Structure of Spacetime\footnote{Expanded version of the lectures given on several occasions including (i) Plenary talk at \textit{Gravity as a Crossroad in Physics}, ERE 2010,  Granada, September  2010; (ii) Keynote address at  \textit{Space-Time-Matter -- Current issues in Quantum Mechanics and beyond}, DICE 2010, Castiglioncello,  September 2010; (iii) Special Lecture at the Indian Academy of Sciences, Bangalore, July 2010; (iv) Plenary talk at \textit{Perspectives in Fundamental Research}, TIFR, March 2010. }}
\author{T. Padmanabhan\\
IUCAA, Pune University Campus, Ganeshkhind,\\
 Pune 411007, INDIA.\\
email:paddy@iucaa.ernet.in }

\date{ }

\begin{document}

\maketitle

\begin{abstract}
I present the theoretical evidence which suggests that gravity is an emergent phenomenon like gas dynamics or elasticity
with the gravitational field equations having the same status as, say, the equations of fluid dynamics/elasticity. This paradigm views  a wide class of gravitational theories ---  including Einstein's theory ---  as describing the thermodynamic limit of the statistical mechanics of `atoms of spacetime'. The evidence for this paradigm is hidden in several classical features of the gravitational theories and depends on just one quantum mechanical input, viz. the  existence of Davies-Unruh temperature of horizons. I discuss several conceptual ingredients of this approach.
\end{abstract}

\section{How can \textit{classical} gravity teach us anything about \textit{quantum} spacetime?}
\label{sec:topdown}

In one sentence, the paradigm that we will explore \cite{rop} is the following: Gravity is an emergent phenomenon like gas dynamics or elasticity
with the gravitational field equations having the same status as, say, the equations of fluid dynamics/elasticity. Historically, this paradigm originated with Sakharov \cite{sakharov} and was interpreted in different ways by Jacobson \cite{ted}, Volovik \cite{grisha}, Bei-Lok Hu \cite{hu} and many others. (Analogue models \cite{analogue} as well as the membrane paradigm \cite{membrane} for black holes have some similarities with this approach. For a  sample of  recent work re-assembling these ideas, see \cite{others}).
I will now elaborate on this theme drawing mainly from the work I was involved in. \footnote{I will use mostly positive signature with English alphabets covering $0,1,...D-1$ and Greek alphabets covering the spatial indices $1,2,...D-1$ of a $D$ dimensional spacetime.}

Part of my programme involves a ``top-down'' approach to quantum spacetime (in the sense of zooming in from the top to smaller and smaller spatial scales, like in a Google map) to learn key lessons (Sections 1--5), which are then used to provide a thermodynamic derivation of field equations from extremising spacetime entropy density (Section 6). Some people use the word ``top-down'' to mean exactly the opposite;  I will use ``top-down'' the way I have defined it, viz. from classical to quantum domain.

One may find it surprising that such an  attempt, to determine the features of the  microscopic theory  from knowing  its properties at the macroscopic scales., is so successful. 
Of course, in the \textit{strict} sense, classical theories cannot tell us anything about quantum dynamics; after all, classical physics, by definition, is independent of  $\hbar$  while quantum effects do depend on $\hbar$. But there is one effect, viz., the thermodynamics of spacetime horizons \cite{daviesunruh} which brings together the principles of quantum theory and gravity. This fact, along with a judicious choice for the questions to ask,  allows one to make a fairly persuasive case for the structure of quantum spacetime. 
To see  such a  ``top-down'' approach in context,  let me describe  at least three other --- more conventional --- examples in which the deeper, more exact, (`bottom layer') theory leaves a tell-tale signature on the `top layer'.

\medskip
\noindent (i)  \textit{Electrons in a helium atom}: Suppose you manage to solve the Schr\"{o}dinger equation for the two electrons in the helium atom and determine the energy eigenfunctions $\psi(\bm x_1, \bm x_2)$. Your experimental friend will tell you that only half of these wave functions [which are antisymmetric under the interchange ($\bm x_1 \Leftrightarrow \bm x_2$)] occur in the real world. Any amount of your staring  at the  Schr\"{o}dinger equation for the helium atom will not tell you why nature requires this antisymmetry under pair exchange for electrons. The reason lies deep down in relativistic quantum theory but its residual effect remains as a tell-tale signature even in the $c=\infty$ limit of field theory, viz., the non-relativistic quantum mechanics.

\medskip
\noindent (ii) \textit{Boltzmann's conjecture of atoms}:
Classical thermodynamics of a gas/fluid uses variables like density, pressure etc. in the continuum description. But the fact that such a fluid can store and exchange  heat energy  \textit{cannot} be understood within the continuum theory. Boltzmann had the insight to suggest that thermal phenomena \textit{demand} the existence of microscopic degrees of freedom in matter. In fact, the law of equipartition, expressed as $E/[(1/2)k_BT] = N$  relates two thermodynamic variables $E, T$ (which are well-defined for a continuum fluid) to $N$,  the number density of \mdof, which cannot  be  interpreted in the  continuum limit at all. The Avogadro's number, closely related to $N$, was determined even before we fully understood what exactly it counts and without any direct evidence for the molecular structure of matter. This is  another example of our being able to say something about microscopic structure from the features of macroscopic theory. 

\medskip
\noindent (iii) \textit{Equality of inertial and gravitational masses}: 
The most dramatic example is provided by Einstein's use of principle of equivalence which, of course, was known for centuries. Einstein realized that $m_i = m_g$  is not a trivial algebraic accident which should be taken for granted, as others before him have done, but requires an explanation. This led him to the description of gravity in terms of the geometry of spacetime. The relation $m_i=m_g$ was a signature of the deeper theory discernible in the approximate (top-layer) description.

The lesson from  these three examples is obvious. For a top down approach to be useful, you need to ask the right questions! One way is to 
 pick up features of the theory that  are usually taken for granted (`algebraic accidents') --- or not even noticed ---  and demand deeper explanations for them. This is the procedure I will follow in this programme to probe the quantum structure of spacetime from known aspects of classical gravity. 

\section{The  conceptual background}

I will describe several peculiar features of classical gravitational theories from which we can obtain a broad picture regarding the quantum microstructure of the spacetime, in the form of  a series of ``lessons''. Most of these (starting from lesson 4!) will deal with specific mathematical features of the theory. But to provide the necessary backdrop, I will distill out of these mathematical features three conceptual points  and  describe them right at the outset, even though the explicit evidence for these will emerge only later on, in the course of the discussion.

\vskip 0.1in
 \subsection*{\textbf{\itshape Lesson 1: Providing a quantum description of spacetime structure is quite different from constructing a  quantum theory of gravity.}} 
\vskip 0.1in
In this approach, it is necessary to  make a clear distinction  between quantum description of spacetime structure and a theory of quantum gravity.  

Classical field equations of gravity \textit{also} happens to describe the classical dynamics of the spacetime because of the geometrical interpretation. In the emergent gravity paradigm, these field equations have a status similar to the equations of fluid mechanics or elasticity. So, if this paradigm is correct,  
 one should \textit{not} expect quantizing a classical theory of gravity to lead us to the quantum structure of spacetime any more than quantizing the equations of elasticity or hydrodynamics will lead us to atomic structure of matter!   Quantizing the elastic vibrations of a solid  will lead  only to phonon physics \cite{hu} just as quantizing a classical theory of gravity will lead to graviton physics. The latter could be quite different from a description of quantum structure of spacetime just as phonon physics is quite different from the physics of the atom. 

\subsection*{\textbf{\itshape Lesson 2: The guiding principle to use for understanding the quantum microstructure of the spacetime should be the thermodynamics of horizons.}}
\vskip 0.1in
Combining the principles of GR and quantum theory  is  not  a technical problem that could be solved just by using sufficiently powerful mathematics. It is  more of a conceptual issue and  decades of failure of sophisticated mathematics in delivering  quantum gravity indicates that we should  try a different approach.
This is very much in tune with item (iii) mentioned in Sec. \ref{sec:topdown}. Einstein did not create a sophisticated mathematical model for $m_i$ and $m_g$ and try to interpret $m_i = m_g$. He used thought experiments to arrive at a conceptual basis in which $m_i = m_g$  can be interpreted in a natural manner so that $m_i=m_g$ will cease to be an algebraic accident.
Once this is done, physics itself led him to the maths that was needed.
 
Of course, the key issue is what could play the role of a guiding principle similar to principle of equivalence in the present context. 
\textit{For this,
my bet will be on the thermodynamics of horizons.}\cite{rop,tpPR}
A successful model will have the connection between horizon thermodynamics  and gravitational dynamics \textit{ at its foundation} rather than this feature appearing as a  result derived in the context of certain specific solutions to the field equations. We will see evidence for its importance throughout the discussion in what follows.

\subsection*{\textbf{\itshape Lesson 3: Think beyond Einstein gravity,  black hole thermodynamics and think off-shell.}}
\vskip 0.1in
There are four technical points closely related to the above conjecture (viz., thermodynamics of horizons should play a foundational role)  which needs to be recognized if this approach has yield dividends:

\begin{itemize}
 
\item  One must concentrate on the general context of observer dependent, local, thermodynamics associated with the local horizons, going beyond the  \textit{black hole} thermodynamics.
Black hole horizons in the classical theory are far too special, on-shell, global constructs to provide a sufficiently general back-drop to understand the quantum structure of spacetime. The preoccupation with the black hole horizons loses sight of the conceptual fact that all horizons are endowed with temperature as perceived by the appropriate class of observers. Observer dependence \cite{tpdialogue} of thermal phenomena is a feature and not a bug!

\item
One should also think beyond  Einstein's theory and use the structure of, say, \LL\ models of gravity \cite{lovelock} in exploring the microstructure of spacetime. Previous work  (starting from ref. \cite{TPParis}) has shown that the interpretation of gravity as an emergent phenomenon transcends Einstein's theory and remains applicable to (at least) all \LL\ models. Exploiting this connection will allow us to discriminate between results of general validity from those which are special to Einstein's theory in $D=4$. Irrespective of  whether \LL\ models are relevant to real world, they provide a good test-bed to see which  concepts and results  are robust and general. 
\item
A corollary is that one should \textit{not} think of entropy of horizons as being proportional to their area. This result, which is true in Einstein's theory, fails for all higher order \LL\ models \cite{wald}. But all the general thermodynamic features  still continue to remain valid. Because area brings in several other closely related geometrical notions, restricting oneself to Einstein's theory leads to an incorrect view of what entropy and quantum microstructure of spacetime could be. 
\item
The quantum features of a theory are off-shell features. But, fortunately,  action principle provides a window to quantum theory because of the path integral formalism. Therefore any peculiar feature of a classical action functional could give us insights into the underlying quantum theory much more than the structure of field equations. This suggests that we need to look at the off-shell structure of the theory using the form of action principles rather than tie ourselves down to field equations.
\end{itemize}

These ingredients, to a great extent, distinguish the approach I was developing from those of many others.

\section{Lessons from the thermodynamics of horizons}

Having outlined the broad conceptual features, I will now move on to specifics. There are four lessons one can learn from putting together well-known features of horizon thermodynamics in an appropriate manner.

\vskip 0.1in
\subsection*{\textbf{\itshape Lesson 4: Temperature of horizons does not depend on the field equations of the theory and is just an indication   that spacetimes, like matter, can be hot, but in a observer-dependent manner.}}
\vskip 0.1in

\noindent
One can associate a temperature with any null surface  that can act as horizon for a class of observers, in any spacetime (including flat spacetime).  This temperature  is determined by the behaviour of the metric close to the horizon and has\textit{ nothing to do with the field equations} (if any) which are obeyed by the metric.  

The simplest situation is that of Rindler observers in flat spacetime with acceleration $\kappa$ who will attribute a temperature 
$  k_BT=(\hbar/c)(\kappa/2\pi)$
 to the Rindler horizon --- which is just a $X=T$ surface in the flat spacetime having no special significance to the inertial observers.  While this result is  usually proved for an eternally accelerating observer, they also hold in the (appropriately) approximate sense for an observer with variable acceleration \cite{dawoodtp10}. In general, this result can be used to show that the vacuum state in a freely falling frame will appear to be a thermal state  in the locally accelerated frame for high frequency modes if $\kappa^{-1}$ is smaller than the local radius of (spacetime) curvature.

In the usual context of  a bifurcation horizon that divides  the spacetime into two causally disconnected regions $R$ and $L$, the global vacuum state $|{\rm vac}\rangle$ of a quantum field theory can be described by a vacuum functional $\langle {\rm vac} | \phi_L, \phi_R\rangle$ in terms of the field configurations $\phi_L$ in $L$ and $\phi_R$ in  $R$. Using the Euclidean path integral representation for the ground state functional, one can express \cite{leeunruh} this functional in two different ways and obtain:
\begin{equation}
\langle {\rm vac} |\phi_L , \phi_R \rangle \propto 
\int_{T_E=0;\phi=(\phi_L,\phi_R)}^{T_E=\infty;\phi=(0,0)}{\cal D}\phi e^{-A}\
\propto
\int_{\kappa t_E=0;\phi=\phi_R}^{\kappa t_E=\pi;\phi=\phi_L}{\cal D}\phi e^{-A}
\propto \langle \phi_L|e^{-(\pi/\kappa )H_R}|\phi_R\rangle
\end{equation} 
where $H_R$ is the  Hamiltonian describing the dynamics in one of the wedges and $\kappa$ is the  acceleration. Both path integrals cover the upper-half of the Euclidean $X-T_E$ plane. The first path integral is in the global coordinate system (inertial, Kruskal ....) with time $T_E$ running from $T_E=0$ to $T_E=\infty$ with the boundary conditions at both limits as indicated. The second path integral is in the coordinate system adapted to region outside the horizon (Rindler, Schwarzschild .....) with the time coordinate behaving like a polar angle in the plane, going from $\kappa t_E=0$ (the right wedge) to $\kappa t_E=\pi$ (the left wedge), with the fields taking appropriate boundary values in the two limits. Thus we get:
\begin{equation}
\langle {\rm vac} |\phi_L , \phi_R \rangle \propto
\langle \phi_L|e^{-(\pi/\kappa )H_R}|\phi_R\rangle
\end{equation} 
For describing the physics in the region outside the horizon, say in $R$, we will trace out the  
 modes $\phi_L$ beyond the horizon. This gives a thermal density matrix for the observables in the right wedge:
\begin{equation}
\rho ( \phi_R',\phi_R)\propto 
\int {\cal D}\phi_L \langle \phi_L , \phi_R'  | {\rm vac}\rangle \langle {\rm vac} |\phi_L , \phi_R \rangle 
\propto
\langle \phi'_R|e^{-(2\pi/\kappa )H_R}|\phi_R\rangle                                                    
\end{equation} 
corresponding to the horizon temperature $T=\kappa/2\pi$.
This result only depends on the near horizon geometry having the approximate form of a Rindler metric 
and is  independent of the field equations of the theory. 

\subsection*{\textbf{\itshape Lesson 5: All thermodynamic variables are observer dependent.}}
\vskip 0.1in
An immediate consequence, not often emphasized, is that \textit{all} thermodynamic variables must become observer dependent if vacuum acquires an observer dependent temperature. A ``normal'' gaseous system with ``normal'' thermodynamic variables ($T, S, F$ etc.....) must be considered as a highly excited state of the inertial vacuum. It is obvious that a Rindler observer will attribute to this highly excited state different thermodynamic variables compared to
what an inertial observer will attribute.
Thus thermal effects in the accelerated frame brings in \cite{tpdialogue,marolf} a new level of observer dependence even to \textit{normal} thermodynamics. One need not panic if variables like entropy now acquire an observer dependence and loses their absolute nature.

\subsection*{\textbf{\itshape Lesson 6: In sharp contrast to temperature, the entropy of horizons depends on the field equations of gravity and cannot be determined by using just QFT in a background metric.}}
\vskip 0.1in
One would have expected that if integrating out certain field modes leads to a thermal density matrix $\rho$, then the entropy of the system should be related to lack of information about the \textit{same} field modes and should be given by $S= - {\rm Tr}\ \rho \ln \rho$.  This entropy,  called entanglement entropy,   (i) is proportional to  area of the horizon and (ii) is divergent without a cut-off \cite{entang-entropy}. Such a divergence makes the result meaningless and thus we cannot  attribute a unique entropy to horizon using just QFT in a background metric.\footnote{In the literature, one often ``regularizes'' the expression for entanglement entropy by introducing a Planck scale cut-off by hand. This procedure lacks justification because of two reasons. First, a free quantum field theory in flat spacetime should not require any cut-off to give meaningful results. Second, in the conventional approach, there is no way a flat spacetime ($G=0$) quantum field theory  will know anything about Planck length. In fact, the divergence of entanglement entropy and the need for a Planck scale cut-off is an indication that there is no such thing as flat spacetime, 
just as there is no such thing as classical, continuum, solid \cite{tpentangle}.}
That is,
while the temperature of the horizon can  be obtained through the study of test-QFT in an external geometry, one cannot understand the entropy of the horizon by the same procedure. 

This is because, unlike the temperature, the \textit{entropy} associated with a horizon in the theory  depends on  the field equations of the theory, which we will briefly review.
Given the principle of equivalence (interpreted as gravity being spacetime geometry) and principle of general covariance, one could still construct a wide class of theories of gravity. For example, if we take the action functional 
\begin{equation}
  A=\int d^Dx \sqrt{-g}\left[L(R^{ab}_{cd}, g^{ab})+L_{\rm matt}(g^{ab},\phi_A)\right]
\end{equation} 
where $L_{\rm matt} $ is the matter Lagrangian (for some matter variables denoted symbolically as $\phi_A$) and vary the metric with appropriate boundary conditions, we will get the field equations (see e.g., chapter 15 of \cite{gravitation}):
\begin{equation}
 \mathcal{G}_{ab}=P_a^{\phantom{a} cde} R_{bcde}  - 2 \nabla^c \nabla^d P_{acdb}- \frac{1}{2} L g_{ab}
 \equiv \mathcal{R}_{ab}- \frac{1}{2} L g_{ab}=\frac{1}{2}T_{ab}
\end{equation} 
where 
$P^{abcd} \equiv (\partial L/\partial R_{abcd})$.
A nice subclass of theories in which the field equations remain  second order in the metric is obtained if we choose $L$ such that 
$\nabla_a P^{abcd}=0$. The most general scalar functionals $L(R^{ab}_{cd}, \gu ij)$  satisfying this condition  are specific polynomials in curvature tensor which lead to  the  \LL\ models \cite{lovelock} with the field equations:
\begin{equation}
P_{ac}^{de} R_{de}^{bc}  - \frac{1}{2} L \delta_{a}^{b}= \mathcal{R}_{a}^{b}- \frac{1}{2m} \mathcal{R}\delta_{a}^{b} =\frac{1}{2}T_{a}^{b}; \quad \mathcal{R}_{a}^{b} \equiv P_{ac}^{de} R_{de}^{bc}; \qquad \mathcal{R} = \mathcal{R}^a_a
\label{scalarpr}
\end{equation} 
The second form of the equation is valid for the $m-$th order \LL\ model for which $\mathcal{R} = R^{abcd} (\partial L/\partial R^{abcd}) = mL$. 
In the simplest context of  $m=1$ we take $ L\propto R=R/16\pi$ (with conventional normalization), leading to $P^{ab}_{cd}=(32\pi)^{-1} (\delta^a_c\delta^b_d-\delta^a_d\delta^b_c)$,
as well as $\mathcal{R}^a_b = R^a_b/16\pi, \mathcal{G}^a_b = G^a_b/16\pi$ and one recovers Einstein's equations.
The structure of the theory is essentially determined by the tensor $P^{ab}_{cd}$ which has the algebraic symmetries of curvature tensor and is divergence-free in all indices. 

In any such generally covariant theory, the infinitesimal coordinate transformation $x^a \to x^a + q^a$ leads
to the conservation  of a Noether current $J^a$ (which depends on $q^a$) given by:
\begin{equation}
J^a \equiv \left(  2\mathcal{G}^{a}_{b} q^b + Lq^a + \delta_{q}v^a \right)
=2\mathcal{R}^{a}_{b} q^b+\delta_{q}v^a; \qquad \nabla_a J^a = 0.
\label{current}
\end{equation}
where $\delta_{q}v^a$ represents the boundary term in the action which arises for the  variation
of the metric in the form $ \delta g^{ab} = ( \nabla^a q^b + \nabla^b q^a$).
Given $\nabla_a J^a = 0$, we can introduce an anti-symmetric tensor $J^{ab}$ by $J^a = \nabla_b J^{ab}$.
For the \LL\ models, one can determine $\delta_q v^a$ and show that
the $J^{ab}$ and $J^a$ can be expressed in the form
\begin{equation}
J^{ab} = 2 P^{abcd} \nabla_c q_d;\qquad  J^a =  2 P^{abcd} \nabla_b \nabla_c q_d
\label{noedef}
\end{equation} 
The field equations of \LL\ models admit black hole solutions (with horizons) in asymptotically flat spacetime. Studying the physical processes occurring in such spacetimes, one can obtain an expression for the entropy of the horizon (called Wald entropy \cite{wald}) which is 
  closely related to the Noether current $J^a$ as follows:
\begin{equation}
S \equiv  \beta\int d^{D-1}\Sigma_{a}\; J^{a}= \beta  \int d^{D-2}\Sigma_{ab}\; J^{ab}
= \frac{1}{4} \int_\mathcal{H}(32\pi\, P^{ab}_{cd})\epsilon_{ab}\epsilon^{dc} d\sigma
\label{noetherint}
\end{equation} 
where $\beta^{-1}=\kappa/2\pi$ is the horizon temperature and $J^a$ is the Noether current for $q^a = \xi^a$ where $\xi^a$ is the Killing vector corresponding to time translation symmetry of the asymptotically static black hole solution. 
 In the final expression the integral is over any surface with $(D-2)$ dimension which is a spacelike cross-section of the Killing horizon on which the norm of $\xi^a$ vanishes,
with $\epsilon_{ab}$ denoting
the  bivector normal to the bifurcation surface.  Thus horizon entropy is given by an integral over the horizon surface of the  $P^{abcd}$, which we may call the \textit{entropy tensor} of the theory.  
Note that the Noether current $J^a$ multiplied by $\beta_{loc}\equiv N\beta$, where $N$ is the lapse function, can be thought of as the entropy current density.

In Einstein's theory, with $32\pi\, P^{ab}_{cd} = (\delta^a_c \delta^b_d - \delta^a_d \delta^b_c)$, 
   the entropy in \eq{noetherint} will be one quarter of the area of the horizon. But in general, the entropy of the horizon is \textit{not} proportional to the area and depends on the theory.\footnote{This feature again shows that the entanglement entropy --- which is always proportional to the horizon area in the conventional approach and is independent of the field equations obeyed by the metric --- cannot be identified with the entropy of the \LL\ models without modifying the regularization procedure. In the emergent paradigm one can argue that such a modification is indeed required. Then, using a generalisation of ideas described in ref.\cite{first},  one can possibly tackle this issue. I will not this discuss here; for more details, see ref. \cite{tpentangle}.}

This dichotomous situation as regards temperature versus entropy is the first indication that the thermodynamics of the horizon, probed by QFT in a external gravitational field, is just the tip of an iceberg. As we will see the emergent paradigm provides a better understanding of these features.

\subsection*{\textbf{\itshape Lesson 7: The connection between horizon entropy and the conserved current arising from the diffeomorphism invariance demands deeper understanding.}}
\vskip 0.1in
Why should a current $J^a$, conserved due to diffeomorphism invariance of the theory, have anything to do with a thermodynamical variable like entropy of horizons in the theory ?

 In the conventional approach, which views $x^a \to x^a + q^a$ as a relabeling of coordinates, this question has no answer. In contrast, if we take the `active' point of view, we notice that $x^a \to x^a + q^a$ also shifts (virtually) the location of null surfaces and thus the information accessible to specific observers. The connection with entropy arises due to the cost of gravitational entropy involved in the virtual displacements of null surfaces.

This idea can be made more precise in terms of entropy balance at local Rindler horizons \cite{entdenspacetime}. Let us choose any event $\mathcal{P}$ and introduce a local inertial frame (LIF) around it with Riemann normal coordinates $X^a=(T,\mathbf {X})$ such that $\mathcal{P}$ has the coordinates $X^a=0$ in the LIF.  
Let $k^a$  be
 a future directed null vector at $\mathcal{P}$ and we align the coordinates of LIF
 such that it lies in the $X-T$ plane at $\mathcal{P}$. We next  transform from the LIF to a local Rindler frame LRF with acceleration $\kappa$ along the $X$ axis.  Let $\xi^a$ be the approximate Killing vector corresponding to translation in the Rindler time such
that the vanishing of $\xi^a\xi_a \equiv -N^2$ characterizes the location of the 
local horizon $\mathcal{H}$ in LRF. Usually, we shall do all the computation 
on a time-like surface infinitesimally away from $\mathcal{H}$
with $N=$ constant,  called a  ``stretched horizon''. 
 Let the time-like unit normal to the stretched horizon
be  $r_a$. 

Consider an infinitesimal displacement of a local patch of the stretched horizon in the direction of $r_a$, by an infinitesimal proper distance $\epsilon$, which will change the proper volume by $dV_{prop}=\epsilon\sqrt{\sigma}d^{D-2}x$ where $\sigma_{ab}$ is the metric in the transverse space.
 The flux of energy through the surface will be  $T^a_b \xi^b r_a$ and the corresponding  entropy flux
 can be obtained by multiplying the energy flux by $\beta_{\rm loc}=N\beta$ which corresponds to the properly redshifted, local, Tolman temperature.  Hence
 the `loss' of matter entropy to the outside observer because the virtual displacement of the horizon has engulfed some matter is 
$
\delta S_m=\beta_{\rm loc}\delta E=\beta_{\rm loc} T^{aj}\xi_a r_j dV_{prop}. 
$
Recalling from \eq{noetherint} that $\beta_{loc}J^a$ gives the gravitational entropy current, the
change in the gravitational entropy is given by $\delta S_{\rm grav} \equiv  \beta_{loc} r_a J^a dV_{prop}$ where $J^a$ is the Noether current corresponding
to the local Killing vector $\xi^a$ given by $J^a=2\mathcal{G}^a_b\xi^b+L\xi^a$. 
As the stretched horizon approaches the true horizon, it can be shown that  $N r^a \to \xi^a$
 and $\beta \xi^a \xi_a L \to 0$. Hence we get, in this limit:
$
\delta S_{\rm grav} \equiv  \beta \xi_a J^a dV_{prop} = 2 \beta \mathcal{G}^{aj}\xi_a \xi_j dV_{prop}.
$
Comparing $\delta S_{\rm grav}$ and $\delta S_m$ we see that the field equations $2\mathcal{G}^a_b=T^a_b$ can be interpreted as the entropy balance condition $\delta S_{grav}=\delta S_{matt}$ thereby providing direct thermodynamic interpretation of the field equations as local entropy balance in local Rindler frame.

 In the emergent paradigm, the spacetime is analogous to a solid made of atoms and $x^a\to x^a+q^a(x)$ is analogous to the deformation of an elastic solid \cite{tpijmp04}. When such a deformation leads to changes in accessible information --- like when one considers the virtual displacements of horizons --- it costs some amount of gravitational entropy thereby providing a direct link between the transformation $x^a\to x^a+q^a(x)$ and spacetime entropy --- a link that is lacking in the conventional approach. We will say more about this in Sec. \ref{sec:entmax}.
 
\section{Thermodynamic interpretation of field equations and action functionals}

I stressed in Sec. \ref{sec:topdown} that for the top-down approach to be of  use, we need to identify the `algebraic accidents' in the top level description which are usually taken for granted without a demand for explanation. I will briefly summarize three such issues in classical gravity, which can give us clues about the microscopic theory. 

\subsection*{\textbf{\itshape Lesson 8: The gravitational field equations reduce to a thermodynamic identity on the horizon in a wide class of theories.}}
\vskip 0.1in

 It can be shown that \cite{tpdawoodgentds} the field equations in any \LL\ model, when evaluated on a static solution of the theory which has a horizon, can be expressed in the form of a thermodynamic identity $TdS = dE_g + PdV$. Here $S$ is the correct Wald entropy of the horizon in the theory, $E_g$ is a geometric expression involving an integral of the scalar curvature of the sub-manifold  of the horizon and $PdV$ represents the work function of the matter source. The differentials $dS, dE_g$ etc. should be thought of as indicating the difference in $S,E_g$ etc between two solutions in which the location of the horizon is   infinitesimally displaced.

This  equality  between field equations on the horizon and the thermodynamic identity --- originally obtained \cite{tdsingr} for spherical horizons in Einstein's theory, 
  has now been demonstrated for an impressively wide class of models \cite{KSP}
like  stationary
axisymmetric horizons and evolving spherically symmetric horizons
in Einstein gravity, static spherically symmetric
horizons and dynamical apparent horizons in
\LL\ gravity, 
generic, static horizon in \LL\ gravity,
 three dimensional BTZ black hole
horizons, FRW cosmological
models in various gravity
theories and even in the case Horava-Lifshitz gravity. 

This result is non-trivial in the sense that the field equation on the horizon does not look very ``thermodynamical''
at first sight. For example, in the simplest context of spherically symmetric horizon in Einstein's theory [with $-g_{00} = g_{11}^{-1} = f(r) $ with $f(a) =0$ determining the location of the horizon at $r=a$], the field equation on the horizon reduces to
\begin{equation}
 \frac{c^4}{G}\left[{\kappa a\over c^2}  - {1\over 2}\right] =  4\pi P a^2
\end{equation} 
where $\kappa = f'(a)/2$ is the surface gravity and $P$ is the pressure of the source. As I said, this equation does not seem to have any thermodynamics in it. However, if we multiply it by $da$ it can be re-written in the form:
\begin{equation}
 \underbrace{\frac{\hbar} {c}\left(\frac{\kappa}{2\pi}\right) }_{\begin{minipage}[c]{2em}
    \vskip 0.1in {$k_BT$}\end{minipage}}
    \ \underbrace{\frac{c^3}{G\hbar}d\left( {1\over 4} 4\pi a^2 \right)}_{\begin{minipage}[c]{1em}
   \vskip 0.1in {$k_B^{-1}dS$}\end{minipage}} 
  \ \underbrace{-\ {1\over 2}\frac{c^4 da}{G}}_{\begin{minipage}[c]{2em}
    \vskip 0.1in {$-dE_g$}\end{minipage}}
 = \underbrace{  P d \left( {4\pi \over 3}  a^3 \right)  }_{\begin{minipage}[c]{2em}
    \vskip 0.1in {$P\, dV$}\end{minipage}}
\end{equation} 
The only extra input we needed was the expression for the horizon  temperature in terms of the surface gravity which needed introducing $\hbar$ in the numerator and denominator. 
Similar miracle occurs in all the gravitational theories, much more general than Einstein's theory, in which entropy is no longer proportional to horizon area.
As we discussed earlier, the temperature of the horizon knows \textit{nothing} about the field equations of the theory but the entropy does. It is therefore remarkable that one obtains the correct combination $TdS$ for a wide variety of theories  showing that the information about the theory is encoded in the entropy functional, exactly as it would be for a macroscopic body.   

There are significant differences between this identity $TdS = dE_g + PdV$ to which field equations reduce to and the so called Clausius relation $TdS = dE_m$ (used, for example, by Jacobson \cite{ted}) which need to be recognised: 
\begin{itemize}
 \item 
 In addition to the obvious existence of the work term $PdV$, it should be stressed that $E_m$ used in the Clausius relation $TdS = dE_m$  is related to \textit{matter  stress tensor} while $E_g$ in the $TdS = dE_g + PdV$ is a purely geometrical construct built out of the metric. The origin of these differences can be traced to two different kinds of virtual displacements of the horizons considered in these two approaches to define the infinitesimal differences \cite{dawoodnew}. 
\item
More importantly, \textit{while $TdS = dE_g + PdV$ holds in widely different contexts,} it has been found to be \textit{impossible} to generalize $TdS = dE_m$ beyond Einstein's theory without introducing additional assumptions (like dissipation), the physical meaning of which remains unclear.

\end{itemize}

Incidentally, while Davies-Unruh temperature scales as $\hbar$ the entropy scales as $1/\hbar$ (coming from inverse Planck area), thereby making $TdS$ independent of $\hbar$! This is reminiscent of the fact that in normal thermodynamics $T\propto 1/k_B,S\propto k_B$ making $TdS$ independent of $k_B$. In both cases, the effects due to discrete microstructure (indicated by non-zero $\hbar$ or $k_B$) disappears in the continuum limit thermodynamics. Thermal phenomena requires microstructure but thermodynamical laws are independent of it!
Similarly we expect the thermodynamic description of spacetime to be  useful and independent of exact nature of the QG description. Any  (`bottom-up``) model for quantum gravity which leads to  horizon thermodynamics and gives Davies-Unruh temperature for QFT in the semi-classical limit, must be consistent with the (`top-down') thermodynamic description merging together in the correct limit.
 
\subsection*{\textbf{\itshape Lesson 9: Holographic structure of gravitational action functionals finds a natural explanation in the thermodynamic interpretation of the field equations.}}
\vskip 0.1in
If the gravitational dynamics and horizon thermodynamics are so closely related, with field equations becoming thermodynamic identities on the horizon, then the action functionals of the theory (from which we obtain the field equations) must contain information about this connection.
This clue comes in the form of another unexplained algebraic accident related to the structure of the action functional
and tells us something significant about the \textit{off-shell structure} of the theory. 

Gravity is the only theory known to us for which the natural action functional preserving symmetries of the theory contain second derivatives of the dynamical variables but still leads to second order differential equations. Usually, this is achieved by separating out the terms involving the second derivatives of the metric into a surface term which is either ignored or its variation is cancelled by a suitable counter-term.  However, this leads to a serious conceptual mystery in the conventional approach
when we recall the following two facts together: (a) The field equations can be obtained by varying the bulk term after ignoring (or by canceling with a counter-term) the surface term. (b) But if we evaluate the surface term on the horizon of any solution to the field equations of the  theory,  one obtains the entropy of the horizon! \textit{How does the surface term, which was discarded before the field equations were obtained, know about the entropy associated with a solution to those field equations?!} In the conventional approach 
we need to accept it as another `algebraic accident' without any
explanation and, in fact,  no explanation is possible within the standard framework. 

The explanation lies in the fact that the surface and bulk term of the Lagrangian are related in a specific manner 
thereby duplicating the information about the horizon entropy \cite{ayan}. One can show that there exists a relation of the form:
\begin{equation}
 \sqrt{-g}L_{\rm sur}=-\partial_a\left(g_{ij}
\frac{\delta \sqrt{-g}L_{\rm bulk}}{\delta(\partial_ag_{ij})}\right)
\label{surbulk}
\end{equation} 
All \LL\ action functionals have this form \cite{TPParis}.  In fact, this relation is crucial for an action with second derivatives of the dynamical variables to still lead to field equations which are only second order --- a
 feature  shared by all the \LL\ models.  It can be shown that this result will be true for  actions that  
can be separated  into a surface term and a bulk term with the surface term being an integral over  $\partial_a
(q^A \pi^a_A)$ where $q^A$ are the dynamical variables and $\pi^a_A$ are the canonical momentum.
This structure allows one to interpret all these action functionals, including Einstein-Hilbert action, as providing the momentum space description (see p. 292 of \cite{gravitation}) of the theory. 

The duplication of information between surface and bulk term in \eq{surbulk} also allows one to obtain the full action \cite{tpPR} from the surface term alone using the entropic interpretation.
In fact, in the the Riemann normal coordinates around any event  $\mathcal{P}$ the gravitational action reduces to a pure surface term, again showing that the dynamical content is actually stored on the boundary rather than in the bulk.

We can also use this fact to relate  the variation of the surface term to $\mathcal{R}^{a}_{b}$ of the theory. From \eq{current}, it follows that:
\begin{eqnarray}
 \int_{\partial\mathcal{V}}d^{D-1}x\sqrt{h} n_a(\delta_qv^a)
&=&\int_{\mathcal{V}}d^{D}x\sqrt{-g} \nabla_a(\delta_qv^a)\nonumber\\
&=&\int_{\mathcal{V}}d^{D}x\sqrt{-g} \nabla_a(2\mathcal{R}^{a}_{b} q^b)
= \int_{\partial\mathcal{V}}d^{D-1}x\sqrt{h} n_a (2\mathcal{R}^{a}_{b} q^b)
\end{eqnarray} 
Computing the corresponding variation of matter action under the change $\delta g^{ab}=\nabla^aq^b+\nabla^bq^a$, one can construct a variational principle to obtain the field equations, purely from the surface term \cite{TPsurfaceaction}. More importantly, since the variation of the surface term gives the change in the gravitational entropy, we see that $\mathcal{R}^{ab}$ essentially determines the gravitational entropy density of the spacetime. We will say more about this in sec. \ref{sec:entmax}.

\subsection*{\textbf{\itshape Lesson 10: Gravitational actions have a surface and bulk terms  because they give the entropy and energy of a static spacetimes with horizons, adding up to make the action the free energy of the spacetime.}}
\vskip 0.1in
This provides yet another, direct, physical interpretation for the structure of the  gravitational action functionals analyzed above. The result is most easily seen for any \LL\ model by writing the time component of the Noether current in \eq{current} for the Killing vector $q^a = \xi^a = (1,\mathbf{0})$ in the form:
\begin{equation}
L = \frac{1}{\sqrt{-g}} \partial_\alpha \left( \sqrt{-g}\, J^{0\alpha}\right) - 2 \mathcal{G}^0_0
\label{strucL}
\end{equation} 
Only spatial derivatives contribute in the first term on the right hand side when the 
spacetime is static. Integrating over $L\sqrt{-g}$ to obtain the action it is is easy to see (using \eq{noetherint}) that the first term gives the entropy and the second term can be interpreted as energy \cite{sanvedtp}.

Finally, I stress again that the real importance of these results arises from the fact that 
they hold  for all \LL\ models  in an identical manner.

\section{The Avogadro number  of the  spacetime}

The results described in the previous sections suggest that there  is a deep connection between horizon thermodynamics and the gravitational dynamics. Because the spacetime can be heated up just like  a body of gas,  the Boltzmann paradigm (``If you can heat it, it has microstructure'') motivates the study of the \mdof\ of the spacetime exactly the way people studied gas dynamics \textit{before} they understood the atomic structure of matter. 
There exists, fortunately, an acid test  of this paradigm which it passes with flying colours. 

\subsection*{\textbf{\itshape Lesson 11: Gravitational field equations imply the law of equipartition $\Delta E=(1/2)k_BT\Delta N$ in any static spacetime, allowing the determination of density of \mdof. The result again displays holographic scaling.}}
 \vskip 0.1in
  Boltzmann's conjecture led to the equipartition law $\Delta E = (1/2) k_BT \Delta N$ relating the number density $\Delta N$ of \mdof\ required to store an energy $\Delta E$ at temperature $T$ and to the determination of Avogadro number of a gas.  If our ideas are correct, we should be able to relate the $E$ and $T$ of a given spacetime to determine the number density of \mdof\ of the spacetime. Remarkably enough, this can be done directly from the field equations \cite{surfaceprd}. In a hot spacetime, Einstein's equations \textit{imply} the equipartition law
\begin{equation}
  E  =  \frac{1}{2}k_B
 \int_{\partial\cal V}
 \frac{\sqrt{\sigma}\, d^2x}{L_P^2}\ 
 \left\{\frac{N a^\mu n_\mu}{2\pi}\right\}
\equiv \frac{1}{2} k_B \int_{\partial\cal V}dn\, T_{\rm loc}
\end{equation}
(where $T_{\rm loc} = (Na^\mu n_\mu /2\pi)$ is the local acceleration temperature and $\Delta n = \sqrt{\sigma}\, d^2 x/ L_P^2$)
thereby allowing us to read off the number density of \mdof.  We again see that gravity is holographic in the sense that  the number density $\Delta n$  scales as the proper area $\sqrt{\sigma}\, d^2 x$  of the boundary of the region rather than the volume. 
(In the case of a gas, we would have got an integral over the volume of the form $dV(dn/dV)$ rather than an area integral.) We also notice that, in Einstein's theory, the number density $(dn/dA)$  is a constant with  every Planck area contributing  a single degree of freedom.

The true elegance of this result again rests on the fact that it holds true for all \LL\ models! For a \LL\ model with an entropy tensor $P^{ab}_{cd}$ one gets the result
\begin{equation}
 E=\frac{1}{2}k_B\int_{\partial\cal V} dn T_{loc}; \qquad
 \frac{dn}{dA}=\frac{dn}{\sqrt{\sigma}d^{D-2}x}=32\pi P^{ab}_{cd}\epsilon_{ab}\epsilon^{cd}
\label{diffeoeqn}
\end{equation} 
where $\epsilon_{ab}$ is the binormal on the codimension-2 cross-section.
All these  gravitational theories are holographic and the density of \mdof\ encodes information about the theory through the entropy tensor. \textit{I  consider these results as the most direct evidence for the emergent paradigm of gravity.}

\subsection*{\textbf{\itshape Lesson 12: One can obtain the Wald entropy for a general theory directly from law of equipartition.}}
\vskip 0.1in

The density of \mdof\ obtained in \eq{diffeoeqn} suggests that the entropy associated with a \textit{general surface} in \LL\ models  (or the entropy associated with a horizon in a more general theory)  will be proportional to an integral over $P^{ab}_{cd}\epsilon_{ab}\epsilon^{cd}$. That is,\footnote{Note that in Einstein's theory, we get $\Delta n=\Delta A/L_P^2$. One usually considers this as arising due to dividing the area $\Delta A$ into $\Delta n$ patches of area $L_P^2$. If we attribute $f$ internal states to each patch, then the total number of microstates $\Delta\Omega$ will be $\Delta\Omega=f^{(\Delta n)}$ and $\Delta S=\ln \Delta\Omega\propto \Delta n$ which is how the extensivity $\Delta S\propto \Delta n$ arises.
In a more general theory, we replace $\Delta n=\Delta A/L_P^2$ by the expression in 
\eq{diffeoeqn}.}
\begin{equation}
S\propto\int_{\partial\mathcal{V}} dn \propto\int_{\partial\mathcal{V}}32\pi P^{ab}_{cd}\epsilon_{ab}\epsilon^{cd}\sqrt{\sigma}d^{D-2}x
\label{waldentro}
\end{equation} 
This is precisely the expression for Wald entropy \cite{wald} \textit{but we have obtained it using only the equipartition law and as a local statement}!

This comes about because the field equations have a specific relationship with Noether current. Further field equations imply equipartition law while Noether current is related to Wald entropy, thereby connecting all the three.
Let me indicate how this comes about by a more direct analysis.
 In static spacetimes, we have a Killing vector $\xi^a$ corresponding to time translation invariance. If we take $q^a=\xi^a$, the expression for the Noether current is quite simple and we get $J^a=2\mathcal{R}^a_b\xi^b$.  Using the relations $J^a \equiv \nabla_b J^{ab}, \xi^a=Nu^a$ and the antisymmetry of $J^{ab}$ one can easily show that:
\begin{equation}
  D_\alpha (J^{b\alpha} u_b)=2N\mathcal{R}_{ab} u^a u^b  
\end{equation} 
This is a generalization of the relation $D_\mu(N a^\mu) = 4\pi \rho_{\rm komar}$ between the divergence of the acceleration and the Komar energy density in Einstein's theory, once again showing the  role of Noether potential $J^{ab}$ in the dynamics.  
The integral version of this relation for a region $\mathcal{V}$ bounded by $\partial\mathcal{V}$ is:
\begin{equation}
  \int_{\partial\mathcal{V}}d^{D-2}x  \sqrt{\sigma}(n_iu_bJ^{bi}) 
= \int_{\partial\mathcal{V}}d^{D-2}x  \sqrt{\sigma} (N n_\alpha J^{\alpha 0})=\int_\mathcal{V} 2 N\mathcal{R}_{ab} u^a u^b \sqrt{h}\, d^{D-1}x 
\label{identity}
\end{equation}
where we have used $u_a=-N\delta_a^0$ and $J^{0\alpha}=-J^{\alpha 0}$. (The middle relation shows that the result is essentially an integral over $\partial\mathcal{V}$ of $J^{bi}d\sigma_{ib}$, where $d\sigma_{ib}=(1/2)n_{[i}u_{b]} \sqrt{\sigma}d^{D-2}x$.)
Now consider a static spacetime with a  
bifurcation horizon $\mathcal{H}$ given by the surface $N^2 \equiv - \xi^a \xi_a =0$. The horizon  temperature $T \equiv\beta^{-1}= \kappa/2\pi$ where $\kappa$ is the surface gravity.
Since the Wald entropy  of the horizon is essentially the Noether charge (multiplied by $\beta$), 
we will interpret  \cite{entdenspacetime} the Noether charge \textit{density}  $\beta J_b u^b $ (multiplied by $\beta$) as the 
 entropy density of the spacetime as perceived by the static observers with four velocity $u^a= \xi^a/N$, so that the total entropy is  
\begin{equation}
S_{\rm grav}[u^i] = \beta\int_\mathcal{V}  J_b u^b \sqrt{h}\, d^{D-1}x 
\label{defs1}
\end{equation} 
Using $J^a = 2 \mathcal{R}^a_b \xi^b$ and \eq{identity}  and integrating the expression over a region bounded by the $N=$ constant surface, it is easy to see that
\begin{equation} 
S=\frac{1}{2}\, \beta E
\end{equation} 
which is a statement of equipartition, first obtained \cite{cqgpap} in 2004 in the form of a relation $E=2TS$ in Einstein's theory and is generalized to all \LL\ models in ref.\cite{surfaceprd}. Further, if we take $\partial\mathcal{V}$ to be the horizon $\mathcal{H}$ and use $\beta T=1$, we get the horizon entropy to be
\begin{equation}
 S=\frac{1}{4}\int_\mathcal{H} dn
 =\frac{1}{4}\int_{\mathcal{H}}32\pi P^{ab}_{cd}\epsilon_{ab}\epsilon^{cd}\sqrt{\sigma}d^{D-2}x
\end{equation} 
which is the standard expression for Wald entropy  in a general theory thereby justifying the choice in \eq{defs1}. This ansatz in \eq{defs1} also fixes the proportionality constant in \eq{waldentro} to be $1/4$. 

Our expressions for entropy and energy in differential form are given by $dE_{\rm komar} = (1/2) T_{\rm loc}(dn/dA) dA$, $dS= (1/4) (dn/dA) dA$. The resulting expression for $TdS$ is essentially equivalent to what we found earlier in the case of first law, $TdS = dE_g + PdV$,  applied to infinitesimal horizon displacements when the differentials appearing in the two expressions are properly related.

\subsection*{\textbf{\itshape Lesson 13: Gravity is intrinsically quantum mechanical at all scales}}
\vskip 0.1in
The holographic nature of gravity which I have alluded to several times shows that area elements play a significant role in the microscopic description of the theory.
This is directly related to the fact that the basic unit of the theory is the Planck \textit{area} $\mathcal{A}_P\equiv (G\hbar/c^3)$.
Only by taking a square root, rather artificially, one obtains the Planck \textit{length}. Classical gravity, in fact, should be described using $\mathcal{A}_P$ rather than using $G$ with Newton's law of gravity written in the form
$F = (\mathcal{A}_P c^3/\hbar) (m_1 m_2/r^2)$.  This has the crucial consequence that one cannot really take $\hbar \to 0$ limit at fixed $\mathcal{A}_P$ and call it classical gravity. Gravity is intrinsically quantum mechanical at all scales \cite{tp2002} because of the microstructure of spacetime.

As an aside, one may mention that, strictly speaking, normal matter is also intrinsically quantum mechanical at all scales due to the atomic structure.  For example, one cannot study classical elasticity, say, by taking the  strict, mathematical, limit $\hbar \to 0$ in a crystal lattice,  because such a limit will also make all the electrons in the atom collapse! What we actually do is to keep $\hbar$ nonzero at subatomic scales, ensuring the atomic stability and take the $\hbar \to 0$ limit for the lattice interactions in the continuum limit
to obtain the laws of elasticity.
We need to do something analogous to obtain classical spacetime from quantum spacetime. 
 
\section{Entropy density  of spacetime an its extremisation}\label{sec:entmax}

So far we have been faithfully following the `top-down' philosophy of starting from known results in classical gravity and obtaining consequences which suggests an alternative paradigm. For example, the results in the last section were obtained by starting from the field equations of the theory, rewriting them in the form of law of equipartition and thus determining the density of \mdof. 

Ultimately, however, we have to start from a microscopic theory and obtain the classical results as a consequence. 
We  know that the thermodynamical behaviour of a normal system can be described by an extremum principle for a suitable  potential (entropy, free energy ...) treated as a functional of appropriate variables (volume, temperature ,....). If our ideas related to gravitational theories are correct, it must be possible to obtain the field equations by extremising a suitably defined thermodynamic potential. The fact that null surfaces block information suggests that this thermodynamic potential should be closely related to null surfaces in the spacetime. 
This expectation turns out to be correct \cite{aseemtp}. 

\subsection*{\textbf{\itshape Lesson 14: Gravitational field equations can be obtained from an alternative, thermodynamic, extremum principle.}}
\vskip 0.1in

Recall that `how gravity tells matter to move' can be determined by demanding the validity of  special relativistic laws for \textit{all} locally inertial observers. Similarly, `how matter curves spacetime' can be determined by demanding that the  a suitable thermodynamic potential of the \mdof\ of the spacetime should be an extremum  \textit{all} local Rindler observers. The physical content of this potential (free energy, entropy, enthalpy ....) will depend on the context but the argument works for any one of them. 
The mathematics involves associating with every null vector field  $n^a(x)$ in the spacetime a thermodynamic potential $\Im(n^a)$ which is quadratic in $n^a$ and given by:
\begin{equation}
\Im[n^a]= \Im_{grav}[n^a]+\Im_{matt}[n^a] \equiv- \left(4P_{ab}^{cd} \D_cn^a\D_dn^b -  T_{ab}n^an^b\right) \,,
\label{ent-func-2}
\end{equation}
where  $P_{ab}^{cd}$ is a tensor having the symmetries of curvature tensor and is divergence-free in all its indices and
$T_{ab}$ is a divergence-free symmetric tensor. (Once we get the field equations we can read off $T_{ab}$ as the matter energy-momentum tensor; the notation anticipates this result). We also know that the $P^{abcd}$ with the assigned properties   can be expressed as  $P_{ab}^{cd}=\partial L/\partial R^{ab}_{cd}$ where $L$ is the \LL\ Lagrangian and $R_{abcd}$ is the curvature tensor \cite{rop}. 
This choice in \eq{ent-func-2} will also ensure that the  equations resulting from the entropy extremisation do not contain any derivative of the metric which is of higher order than second. (More general possibilities exist which I will not discuss here.). We now demand that $\delta \Im/\delta n^a=0$ for the variation of all null vectors $n^a$ with the  condition $n_an^a=0$ imposed by adding a  Lagrange multiplier function $\lambda(x)g_{ab}n^an^b$ to $\Im[n^a]$. Using 
\begin{equation}
 \frac{\partial \Im}{\partial ( \nabla_c n^a)}= (-8 P^{cd}_{ab} \nabla_d n^b);\quad
  \frac{\partial \Im}{ \partial  n^a}=2[T_{ab}+\lambda(x)g_{ab}]n^b
\label{sgravder1}
\end{equation}  
the Euler-Lagrange equations reduce to: 
\begin{equation}
\nabla_c\left[ -8 P^{cd}_{ab} \nabla_d n^b\right]=2[T_{ab}+\lambda(x)g_{ab}]n^b
\end{equation} 
Because of the condition $\nabla_cP^{cd}_{ab}=0$ and the antisymmetry $P^{cd}_{ab}=-P^{dc}_{ab}$ we find that all the derivatives disappear on the left hand side and an elementary calculation gives:
\begin{equation}
\left(2\mathcal{R}^a_b -  T{}^a_b-\lambda \delta^a_b\right) n_a=0\,, 
\label{ent-func-6}
\end{equation}
where $\mathcal{R}^a_b\equiv P_{bi}^{jk}R^{ai}_{jk}$. We 
demand that \eq{ent-func-6} should hold for all 
null vector fields
$ n^a$. 
Using the
generalized Bianchi identity and the condition $\nabla_aT^a_b=0$ we obtain \cite{rop,aseemtp} from \eq{ent-func-6} the  equations
\begin{equation}
\mathcal{G}^a_b =  \mathcal{R}^a_b-\frac{1}{2}\delta^a_b L = \frac{1}{2}T{}_b^a +\Lambda\delta^a_b   
\label{ent-func-71}
\end{equation}
where $\Lambda$ is a constant.  These are  precisely the field equations for  gravity  in a theory with \LL\ Lagrangian $L$ (with an undetermined cosmological constant $\Lambda $ which arises as an integration constant. 

The thermodynamical potential can be obtained by integrating the density $\Im[n^a]$ over a region of space or a surface etc. depending on the context. The matter part of the $\Im$ is proportional to $T_{ab}n^an^b$ which will pick out the contribution $(\rho+p)$ for an ideal fluid, which is the enthalpy density. If multiplied by $\beta=1/T$, this reduces to the entropy density because of Gibbs-Duhem relation. When the multiplication by $\beta$ can be reinterpreted in terms of integration over $(0,\beta)$ of the time coordinate, the corresponding potential can be interpreted as entropy and the integral over space coordinates can be interpreted as rate of generation of entropy. [This was the interpretation provided in the earlier works \cite{rop,aseemtp} but the result is independent of this interpretation as long as suitable boundary conditions can be imposed]. One can also think of $\Im[n^a]$ as an effective Lagrangian for a set of collective variables $n^a$ describing the deformations of null surfaces.

In addition to providing a purely thermodynamic extremum principle for the field equations of gravity, the above approach also has the following attractive features.

\begin{itemize}

 \item The extremum value of the thermodynamic potential, when computed on-shell for a solution with static horizon, leads to the Wald entropy. This is a non-trivial consistency check on the approach because it was not designed to reproduce the Wald entropy. It also shows that when the field equations hold, the total entropy of a region $\mathcal{V}$ resides on its boundary $\partial\mathcal{V}$ which is yet another illustration of the holographic nature of gravity.

\item In the semi-classical limit, one can show \cite{entropyquant} that the gravitational (Wald) entropy is quantized with $S_{\rm grav}$ [on-shell] $=2\pi n$. In the lowest order \LL\ theory, the entropy is proportional to area and this result leads to area quantization. More generally, it is the gravitational entropy that is quantized.  The law of equipartition for the surface degrees of freedom is closely related to this entropy quantization because both arise from the existence of discrete structures on the surfaces in question. 

\item The  entropy functional in \eq{ent-func-2}  is invariant under the shift $T_{ab} \to T_{ab} + \rho_0 \gl ab$ which shifts the zero of the energy density. This symmetry allows any low energy cosmological constant, appearing as a parameter in the variational principle, to be gauged away thereby alleviating the cosmological constant problem to a great extent \cite{tpcc}. I will not discuss this issue here.

\end{itemize}

There is another way of interpreting \eq{ent-func-6}  which  is more in tune with the  emergent perspective of gravity. Note that, while \eq{ent-func-6} holds for any vector field once the normalization condition is imposed through the Lagrange multiplier, the entropy was originally attributed to \textit{null} vectors and hence it is natural to study  \eq{ent-func-6} when $n^a = \ell^a$,    the null normal  of a null surface $\mathcal{S}$ in the spacetime and project 
\eq{ent-func-6} onto the null surface.
If $\el$ is the normal to $\mathcal{S}$, then such a projection leads to the equations:
\begin{equation}
R_{mn}\ell^m q^n_{a}=8\pi T_{mn}\ell^m q^n_{a}; \quad R_{mn}\ell^m \ell^n=8\pi T_{mn}\ell^m \ell^n
\label{albertpair}
\end{equation}
where
$q_{ab} = \gl ab+ \ell_a k_b + \ell_b k_a$ with $k^a$ being another auxiliary null vector satisfying $\el \w \cdot \w k = -1$. The metric $q_{ab}$ with $q_{ab} \ell^b =0 = q_{ab} k^b$ acts as a  projector to $\mathcal{S}$ (see ref. \cite{dns} for details).
It is possible to rewrite the first equation in \eq{albertpair} in the form of a Navier-Stokes equation thereby providing a hydrodynamic analogy for gravity.  This equation,  known in the literature as Damour-Navier-Stokes (DNS) equation \cite{damourthesis},  is usually derived by rewriting the field equations. Our analysis \cite{dns} provides an entropy extremisation principle for the DNS equation which makes the hydrodynamic analogy natural and direct. 

It may also be noted that the gravitational entropy density  --- which is the integrand $\Im_{grav}\propto ( -P_{ab}^{cd} \D_c\ell^a\D_d\ell^b)$ in \eq{ent-func-2} --- obeys the relation:
\begin{equation}
 \frac{\partial \Im_{\rm grav}}{\partial ( \nabla_c \ell^a)}\propto (- P^{cd}_{ab} \nabla_d \ell^b) \propto (\nabla_a \ell^c  - \delta^c_a \nabla_i \ell^i)
\label{sgravder}
\end{equation} 
where the second relation is for Einstein's theory. This term is analogous to the more familiar object $t^c_a = K^c_a - \delta^c_a K$ (where $K_{ab}$ is the extrinsic curvature) that arises in the (1+3) separation of Einstein's equations. (More precisely, the  projection to 3-space leads to $t^c_a$.) This combination can be interpreted as a surface energy momentum tensor in the context of membrane paradigm \cite{pricethorn} because $t_{ab}$ couples to $\delta h^{ab}$ on the boundary surface when we vary the gravitational action ( see, e.g., eq.(12.109) of \cite{gravitation}).  Equation~(\ref{sgravder}) shows that  the entropy density of spacetime is directly related to  $t^c_a$ and its counterpart in the case of null surface.   
This term also has the interpretation as the canonical momentum conjugate to the spatial metric in (1+3) context and \eq{sgravder} shows that the entropy density leads to a similar structure. 
That is, the canonical momentum conjugate  to metric in the conventional approach and the momentum conjugate to $\ell^a$ in $S_{\rm grav}$ are essentially the same.

Further, the \textit{functional} derivative  of the gravitational entropy in \eq{ent-func-2} has the form, in any \LL\ model:
\begin{equation}
 \frac{\delta \Im_{\rm grav}}{\delta \ell^a} \propto \mathcal{R}_{ab}\ell^b \propto J_a
\end{equation} 
Previous discussion has shown that the current $J_a= 2\mathcal{R}_{ab} \ell^b$ plays a crucial role in interpreting gravitational field equations as entropy balance equations. In the context of local Rindler frames, when $\ell^a$ arises as  a limit of the time-like Killing vector in the local Rindler frame, $J_a$ can be interpreted as the Noether (entropy) current associated with the null surface. In that case, the generalization of the two projected equations in \eq{albertpair} to \LL\ models will read as
\begin{equation}
 J_a \ell^a = \frac{1}{2} T_{ab} \ell^a \ell^b; \quad J_aq^a_m = \frac{1}{2}T_{ab}\ell^a q^b_m
\end{equation} 
which relate the gravitational entropy density and flux to matter energy density and momentum flux. (The second equation in the above set becomes the DNS equation in the context of Einstein's theory.) All these results, including the DNS equation,  will have direct generalization to \LL\ models which can be structured using the above concepts.
We again see that all these ideas find a natural home in the emergent paradigm. 

\section{Concluding Comments}

As promised, I have presented the internal evidence hidden in the structure of classical gravitational theories which suggest that gravity is an emergent phenomenon. This evidence brings up the holographic nature of gravity in more than one way (surface density of \mdof, structure of gravitational action functionals .....), provides a thermodynamic interpretation to field equations (field equations reducing to $TdS=dE_g+PdV$ on the horizons, entropy balance for virtual displacements of horizons, equipartition ), allows one to explicitly determine the number density of \mdof and --- finally --- derive the field equations from an entropy maximization procedure.  The approach also clarifies several issues which have no explanation in conventional procedure and links several ideas together (like e.g. the relation between the diffeomorphism invariance and the entropy of null surfaces). All of these work in any \LL\ model seamlessly without we having to tinker anything.

It is worthwhile to list explicitly the questions which have natural answers in the emergent paradigm while have to be treated as algebraic accidents in the conventional approach:
\begin{enumerate}

\item While the temperature of the horizon can be obtained using QFT in curved spacetime,  the \textit{corresponding} entanglement entropy is divergent and meaningless. Why?

\item The temperature of horizon is independent of the field equations of gravity but the entropy of the horizon depends explicitly on the field equations. What does this difference signify?

\item The horizon entropy can be expressed in terms of the Noether current which is conserved due to diffeomorphism invariance. Why should an infinitesimal coordinate transformation $x^a \to x^a + q^a$ have anything to do with a thermodynamic variable like entropy?

\item Why does gravitational field equations (which does not look very ``thermodynamical''!) reduce to $TdS = dE_g + PdV$ on the horizon, picking up the correct expression for $S$ for a wide class of theories?

\item How come all gravitational action principle have a surface and bulk term which are related in a specific manner (see \eq{surbulk})? Why do the surface and the bulk terms allow the interpretation as entropy and energy in static spacetimes? 

\item The field equations for gravity can be obtained from the bulk part of the action after discarding the surface term. But the surface term evaluated on the horizon of a solution gives the entropy of the horizon! How does the surface term --- which was discarded before the field equations were obtained --- know about the entropy of a solution?

\item 
 Why does the gravitational field equations reduce to the equipartition form, expressible as $\Delta E = (1/2) (k_B T) \Delta n$ allowing us to determine the analog of Avogadro's number for the spacetime? And, why does the relevant \mdof\ for a region reside on the boundary of the region?

\item Finally, why is it possible to derive the field equations of any diffeomorphism invariant theory of gravity by extremizing an entropy functional associated with the null surfaces in the spacetime, without treating the metric as a dynamical variable?

\end{enumerate}

Obviously, any alternative perspective, including the conventional approach, need to provide the answers for the above questions if they have to be considered a viable alternative to emergent paradigm.  The explanations need to work for
all \LL\ models and not for just Einstein's theory. I think the emergent paradigm scores on all these counts and provides valuable insights into the deeper structure of the theory.


\begin{thebibliography}{99}


\bibitem{rop}
Padmanabhan T 2010  \textit{Rep. Prog. Phys.}  \textbf{73}  046901 [arXiv:0911.5004)].


\bibitem{sakharov}
Sakharov A D  1968  \textit{Sov. Phys. Dokl.} {\bf 12} 1040.


\bibitem{ted}
Jacobson T  1995 \textit{Phys. Rev. Lett.} \textbf{75}  1260.


\bibitem{grisha}
Volovik G E  2003 \textit{The universe in a helium droplet} (Oxford University Press).


\bibitem{hu}
Hu B L 2010 Gravity and Nonequilibrium Thermodynamics of Classical Matter [arXiv:1010.5837]; 
General Relativity as Geometro-Hydrodynamics [arXiv:gr-qc/9607070].



\bibitem{analogue}
For a review, see e.g., Barcelo C,  Liberati S and Visser M 2005
 \textit{Living Rev.Rel.} \textbf{8} No. 12 [gr-qc/0505065];



\bibitem{membrane}
Thorne K S et al. 1986 \textit{Black Holes: The Membrane Paradigm} (Yale University Press)
 

\bibitem{others}
Yu Tian Xiao-Ning Wu arXiv:1012.0411;
 Hendi S H and Sheykhi A  arXiv:1012.0381;
H. Culetu, arXiv:1011.3343;
Wei Gu, Miao Li, Rong-Xin Miao, arXiv:1011.3419;
Li-Ming Cao, arXiv:1009.4540;
Shao-Feng Wu et al. arXiv:1008.2072;
Banerjee R et al 2010 \textit{Phys. Rev.} \textbf{D 82} 124002;
Yu Tian et al., arXiv:1007.4331;
Shalyt-Margolin A.E.  arXiv:1006.4979;
P. Nicolini, arXiv:1005.2996;
F. Piazza, arXiv:1005.5151;
Fursaev D V  2010 \textit{Phys.Rev}.\textbf{D82}:064013;
Kiselev V V et al,  (2010) \textit{Mod.Phys.Lett}.\textbf{A25}, 2223;
Sheykhi A  2010 \textit{Phys.Rev}, \textbf{D81} 104011;
Xiao-Gang He and Bo-Qiang Ma arXiv:1003.1625;
Jae-Weon Lee arXiv:1003.1878; arXiv:1003.4464 ;
Banerjee R and Bibhas Ranjan Majhi, arXiv:1003.2312;
Xiao-Gang He and Bo-Qiang Ma arXiv:1003.2510;
Yi-Fu Cai, Jie Liu and Hong Li arXiv:1003.4526;
Ghosh S arXiv:1003.0285;
Munkhammar J arXiv:1003.1262;
Zhao L arXiv:1002.0488;
 Kowalski-Glikman J arXiv:1002.1035;     
 Yu-Xiao Liu, Yong-Qiang Wang and Shao-Wen Wei,  arXiv:1002.1062;
 Rong-Gen Cai, Li-Ming Cao and Nobuyoshi Ohta, arXiv:1002.1136;
Pesci P arXiv:1002.1257;
 Yu Tian and Xiaoning Wu arXiv:1002.1275;
 Yun Soo Myung and Yong-Wan Kim arXiv:1002.2292;
Vancea  I V and  Santos M A  arXiv:1002.2454;
Konoplya R A arXiv:1002.2818;
Hristu Culetu arXiv:1002.3876;
Rong-Gen Cai, Li-Ming Cao and Nobuyoshi Ohta 2010 \textit{Phys.Rev.} \textbf{D 81} 061501  [arXiv:1001.3470];
Smolin L, arXiv:1001.3668;
Fu-Wen Shu and Yungui Gong arXiv:1001.3237;
Padmanabhan T, arXiv:1001.3380;
Makea J arXiv:1001.3808;
Miao Li and Yi Wang arXiv:1001.4466;
Changjun Gao  arXiv:1001.4585;
Yi Zhang, Yun-gui Gong and Zong-Hong Zhu arXiv:1001.4677;
Hristu Culetu arXiv:1001.4740;
Yi Wang arXiv:1001.4786;
Tower Wang arXiv:1001.4965;
Shao-Wen Wei, Yu-Xiao Liu and Yong-Qiang Wang arXiv:1001.5238;  
Yi Ling and Jian-Pin Wu arXiv:1001.5324;
Jae-Weon Lee, Hyeong-Chan Kim, and Jungjai Lee arXiv:1001.5445;
Verlinde E P arXiv:1001.0785.



\bibitem{daviesunruh}
Bekenstein J D  1972  \textit{Nuovo Cim. Lett.} \textbf{4} 737--740; 
Hawking S W  1975 \textit{Commun. Math. Phys.} \textbf{43}   199--220.;
Davies P C W 1975  \textit{J. Phys.} A \textbf{8}  609--616; 
Unruh W G  1976 \textit{Phys. Rev.} D \textbf{14}  870.



\bibitem{tpPR}
Padmanabhan T    2005 \textit{Phys. Reports}  \textbf{406} 49 [gr-qc/0311036]. 



\bibitem{tpdialogue} 
Padmanabhan T 2009 \textit{A Dialogue on the Nature of Gravity} as Chapter 2 in ``Foundations of Space and Time'', edited by J. Murugan, A. Weltman and G.F.R. Ellis (Cambridge University Press, 2011) [arXiv:0910.0839]. 



 \bibitem{lovelock} 
Lanczos C  1932 {\it Z. Phys.} 
            {\bf 73} 147; 
Lovelock D  1971 {\it J. Math. Phys.} 
            {\bf 12} 498. 


\bibitem{TPParis}
 Padmanabhan T 2005 \textit{Dark Energy: Mystery of the Millennium}  Albert Einstein Century International Conference Paris  \textit{ AIP Conference Proceedings} \textbf{ 861} 858 [astro-ph/0603114]. 



\bibitem{wald} 
Wald R M 1993 \textit{Phys. Rev. D}   {\bf  48}  3427 [gr-qc/9307038];
Iyer V  and  Wald R M 1995 \textit{Phys. Rev. D} {\bf 52}  4430 [gr-qc/9503052].



\bibitem{dawoodtp10}
Kothawala D  and  Padmanabhan T 2010 \textit{ Phys.Letts.}  \textbf{B 690}  201-206  [arXiv:0911.1017];
Raval A, Hu B L and Don Koks  1997 \textit{Phys.Rev} \textbf{D55} 4795 [gr-qc/9606074]. 



\bibitem{leeunruh}
Lee T D 1986 \textit{ Nucl. Phys.} B \textbf{264}  437;
Unruh W G  and Weiss N  1984 \textit{Phys. Rev.} D \textbf{29}  1656;
for a textbook description, see section 14.5 of ref.\cite{gravitation}.



\bibitem{marolf} 
Marolf D,  Minic D and  Ross S 2004 \textit{ Phys.Rev.}  \textbf{D69}  064006.



\bibitem{entang-entropy} See, for e.g., 
Bombelli L et al.  1986 \textit{Phys.Rev.} \textbf{D34} 373;
Srednicki M  (1993) \textit{Phys.Rev.} \textbf{D71} 66 and for a recent review
T. Nishioka \textit{et al}. 2009 \textit{J.Phys.} \textbf{A42} 504008 
	[arXiv:0905.0932].


\bibitem{tpentangle} 
Padmanabhan T  2010 \textit{Phys. Rev. D} \textbf{82} 124025
[arXiv:1007.5066].



\bibitem{gravitation}
Padmanabhan T 2010 \textit{Gravitation: Foundations and Frontiers}, (Cambridge University Press UK).



\bibitem{first} 
Padmanabhan T  1997 \textit{Phys. Rev. Letts} \textbf{78} 1854 [hep-th-9608182];
                   \textit{Phys. Rev.} \textbf{D57} (1998)  6206 ;
               \textit{Class. Quan. Grav.}, \textbf{4}, (1987) L107;
                \textit{ Annals Phys.,} \textbf{165}, (1985) 38-58;
Srinivasan K et al.  1998 \textit{Phys. Rev.} \textbf{D  58} 044009 [gr-qc-9710104].


\bibitem{entdenspacetime}
T. Padmanabhan , \textit{A Physical Interpretation of Gravitational Field Equations}, Plenary talk given at the International Conference on `Invisible Universe', 29 June- 3 July, 2009 Paris; AIP Conference Proceedings, \textbf{ 1241},  93-108 (2010) [arXiv:0911.1403]; 
 Padmanabhan T  2009 \textit{Int.Jour.Mod.Phys.} \textbf{D18 } 2189 [arXiv:0903.1254].
 

\bibitem{tpijmp04}
Padmanabhan T 2004 \textit{Int.Jour.Mod.Phys.}  \textbf{D13} 2293-2298 [gr-qc/0408051]; 
  2005 \textit{Brazilian Jour.Phys.} (Special Issue)  \textbf{35}, 362 [gr-qc/0412068]. 



\bibitem{tpdawoodgentds} 
Kothawala D and Padmanabhan T 2009 \textit{Phys. Rev.}  \textbf{D79} 104020 [arXiv:0904.0215]. 




\bibitem{tdsingr}
 Padmanabhan T 2002 {\it Class. Quan. Grav.} {\bf 19} 5387  [gr-qc/0204019].





\bibitem{KSP} There is large literature on this subject a small sample of which will be:
Cai R G \etal 2008 \textit{Phys. Rev.}  D {\bf 78} 124012;
Kothawala D, Sarkar  S and Padmanabhan T 2007, \textit{Phys. Lett.}  B {\bf 652} 338   [gr-qc/0701002];
Paranjape A, Sarkar S and Padmanabhan T 2006 \textit{Phys. Rev.} D {\bf 74} 104015 [hep-th/0607240];
Akbar M 2007 \textit{Chin. Phys. Lett.}  {\bf 24} 1158  [hep-th/0702029];
Cai R G and Kim  S P  2005 \textit{JHEP} {\bf 0502} 050  [hep-th/0501055];
Sheykhi A,  Wang B and Cai R G  2007 \textit{Nucl. Phys.}  B {\bf 779} 1 [hep-th/0701198];
Sheykhi Wang A B and Cai R G 2007 \textit{Phys. Rev.} D {\bf 76} 023515  [hep-th/0701261];
A. E.Shalyt-Margolin, [arXiv:1006.4979];
Cai  R G 2008  \textit{Prog. Theor. Phys. Suppl.} {\bf 172} 100 [arXiv:0712.2142];
Ge  X H  2007 \textit{Phys. Lett.}  B {\bf 651} 49  [hep-th/0703253];
Gong Y and Wang A 2007 \textit{Phys. Rev. Lett.} {\bf 99} 211301   [arXiv:0704.0793].
Wu S F, Wang B and Yang G H 2008 \textit{Nucl. Phys.}  B {\bf 799} 330 [arXiv:0711.1209];
Wu S F \etal 2008 \textit{Class. Quant. Grav.} \textbf{25} 235018 [arXiv:0801.2688];
Cai R G and Ohta  N  2009 [arXiv:0910.2307].



\bibitem{dawoodnew}
Kothawala D, \textit{The thermodynamic structure of Einstein tensor}, [arXiv:1010.2207].



\bibitem{ayan}
 Mukhopadhyay A and   Padmanabhan T  2006 \textit{Phys.Rev.} \textbf{ D 74} 124023 [hep-th/0608120].


\bibitem{TPsurfaceaction} 
Padmanabhan T  2006 \textit{Gen.Rel.Grav.} \textbf{38} 1547-1552;
  \textit{ Int.J.Mod.Phys.} \textbf{D 15}  2029 (2006) [gr-qc/0609012]. 



\bibitem{sanvedtp} 
Sanved K and  Padmanabhan T 2010  \textit{Phys.Rev.}  \textbf{ D 82} 024036  [arXiv:1005.0619].


\bibitem{surfaceprd}
Padmanabhan T 2010 \textit{Mod. Phys. Lett.} \textbf{A 25} 1129  [arXiv:0912.3165];
 \textit{Phys. Rev.} \textbf{D 81} 124040 (2010) [arXiv:1003.5665].



\bibitem{cqgpap}
 Padmanabhan T  2004 \textit{Class.Quan.Grav.}, \textbf{21}, 4485 [gr-qc/0308070] 



\bibitem{tp2002} 
 Padmanabhan T  2002 \textit{ Mod.Phys.Letts. A}  \textbf{17} 1147 [hep-th/0205278];
  \textit{Gen.Rel.Grav.}  \textbf{34}  2029-2035 (2002) [gr-qc/0205090].



\bibitem{aseemtp} 
Padmanabhan T   2008 \textit{Gen.Rel.Grav.}   \textbf{40} 529-564 [arXiv:0705.2533];
Padmanabhan T   2008 \textit{Gen. Rel. Grav.}  \textbf{40}  2031-2036;
Padmanabhan T  and Paranjape A 2007 \textit{Phys.Rev.} D \textbf{75} 064004 [gr-qc/0701003].



\bibitem{entropyquant}
Kothawala D,  Padmanabhan T and  Sarkar S 2008 \textit{Phys.Rev}. \textbf{D78} 104018  [arXiv:0807.1481]




\bibitem{tpcc}
 Padmanabhan T, \textit{Adv. Sci. Lett.} \textbf{2} 174 (2009) [arXiv:0807.2356];
               \textit{Phys. Rep.} \textbf{380}, 235 (2003) [hep-th/0212290];   
\textit{Class.Quan.Grav.}, \textbf{22}, L107-L110, (2005) [hep-th/0406060].
 


\bibitem{dns}
Padmanabhan T \textit{Entropy density of spacetime and the Navier-Stokes fluid dynamics of null surfaces}, [arXiv:1012.0119].




\bibitem{damourthesis} 
 Damour T 1979 
Th\`ese de doctorat d'\'Etat, Universit\'e Paris  (available at http://www.ihes.fr/$\sim$damour/Articles/).



\bibitem{pricethorn}
Price R H and Thorne, K S 1986 
\textit{Phys. Rev.}  {\bf D 33} 915. 


\end{thebibliography}
\end{document}